# Sarin and Air Permeation Through a Nanoporous Graphene

Marco A. Maria[1,2,3] and Alexandre F. Fonseca[3]

[1] Federal University of São Carlos – Sorocaba.

[2] Facens University Center – Sorocaba.

[3] Applied Physics Department, Institute of Physics "Gleb Wataghin", University of Campinas, Campinas, SP, 13083-970, Brazil.

*Sarin gas is a dangerous chemical warfare agent (CWA). It is a nerve agent capable of bringing a person to death in about 15 minutes. A lethal concentration of sarin molecules in air is about 30 mg/m$^3$. Experimental research on this gas requires very careful safety protocols for handling and storage. Therefore, theoretical and computational studies on sarin gas are very welcome and might provide important safe guides towards the management of this lethal substance. In this work, we investigated the interactions between sarin, air and nanoporous graphene, using tools of classical molecular dynamics simulations. Aiming to cast some light in the possible sarin selective filtration by graphene, we designed a bipartite simulation box with a porous graphene nanosheet placed at the middle. Sarin and air molecules were initially placed only on one side of the box so as to create an initial pressure towards the passage of both to the other side. The box dimensions were chosen so that the hole in the graphene was the only possible way through which sarin and air molecules can get to the other side of the box. The number of molecules that passed through the hole in graphene was monitored during 10 ns of simulation and the results for different temperatures were compared. The results show that, as far as the size of the holes are small, van der Waals forces between graphene and the molecules play a significant role on keeping sarin near graphene, at room temperature.*

## INTRODUCTION

Sarin is an organophosphorus nerve agent that has been used as a chemical warfare. It interacts with nervous system blocking the action of acetylcholinesterase enzyme, what can result in gastroenteritis, weakness, tachycardia and bronchospasms, eventually causing death [1-2].

Due to the dangerousness of sarin and other chemical warfare agents (CWAs), the focus of the most of the research on those substances has been limited to their detection and/or degradation, including its simulants [3-12]. Theoretical methods are thus very welcome to study the chemical and physical properties of CWAs at different physical conditions. Our proposal in this work is, then, to computationally investigate, under several conditions of temperature and gas concentration, the permeation of sarin molecules through a nanoporous graphene, observing the mechanisms for their filtration. Particularly, we are interested in selectively block sarin molecules and allow the passage of air molecules through the holes of a nanoporous graphene. The computational method, the system model and the results are presented in the next sections.

**MODEL AND SIMULATION DETAILS**

The porous graphene structure is shown in Figure 1(a). The three $CH_3$ of the sarin molecule were modelled as unique artificial atoms with net physical properties (mass and charge). These unities were named as C3s and CH3s as shown in Figure 1(b). For the air, it was considered the composition of 78% of $N_2$ (nitrogen), 21% of $O_2$ (oxygen) and 1% of Ar (argon) [13].

In order to study the passage of sarin and air molecules through nanoporous graphene, we consider a simulation box of dimensions $X = 34$ Å, $Y = 32$ Å and $Z = 200$ Å, with the nanoporous graphene sheet fixed at the middle of the box with its plane parallel to the *xy* plane (see Figure 2). Sarin and air molecules are initially placed only in one side of the box, while the other side of box is initially empty, such that the only possible way of a sarin molecule to go from one side to the other of the box is passing through the hole in the graphene. Also, the concentration of sarin and air molecules on one side of the box naturally creates a pressure towards the passage through the porous graphene. Different amounts of sarin concentrations are considered in this study.

LAMMPS package [14] was used to integrate the equations of movement for all atoms of the system according to the following potential energy, $U_{system}$:

$$U_{system} = k_{bond}(r - r_0)^2 + k_{angle}(\theta - \theta_0)^2 + U_{dihedral} + U_E + U_{vdW} \quad (1)$$

where $k_{bond}$ and $k_{angle}$ are the elastic constants related to changes in bond length and bond angle, respectively. $r$ and $\theta$ are the bond distance and bond angles, respectively, and the 0 subscript in eq. (1) is for the equilibrium values of these two quantities. $U_{dihedral}$ is the dihedral term of the potential that is described as follows. For sarin, the expression $U_{dihedral} = \sum k_i [1.0 + \cos(n_i \phi - d_i)]$ is used for dihedral angles formed by the following sequences of atoms: $C1s - O1s - Ps - O2s$, $C3s - C1s - O1s - Ps$ and $C1s - O1s - Ps - CH3s$; and the expression $U_{dihedral} = \sum A_n \cos^{n-1}(\phi)$ is used for the dihedral angles of the following sequence of atoms: $C1s - O1s - Ps - Fs$. The parameters of eq. (1) for sarin and air were taken from references [15] and [16]. $U_E$ and $U_{vdW}$ are the Coulomb and van der Waals terms, respectively.

Before running the simulations on the sarin-air-graphene system, the parameters for sarin and air were tested for certain properties. Using a NPT ensemble, with pressure fixed at 1 atm and temperatures of 298 K for sarin and 330 K for air, we reached the following values of densities: 1,1 g/cm³ for sarin and 0,0014 g/cm³ for air. These results are in good agreement with those presented in the references [15] and [16].

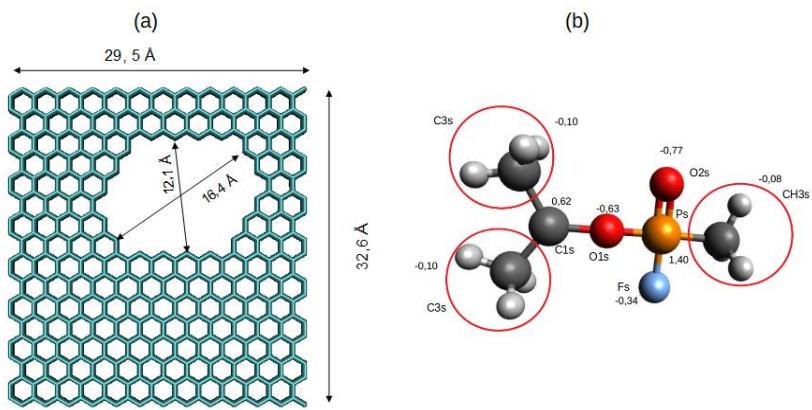

Figure 1. Models of (a) porous graphene and (b) sarin structures. Red circles in sarin indicates the groups of atoms that are considered as one (united-atom). Atom charges and names are shown near the respective atoms. Gray, white, red, orange and blue colours, for the sarin molecule, represent carbon, hydrogen, oxygen, phosphorus and fluorine atoms.

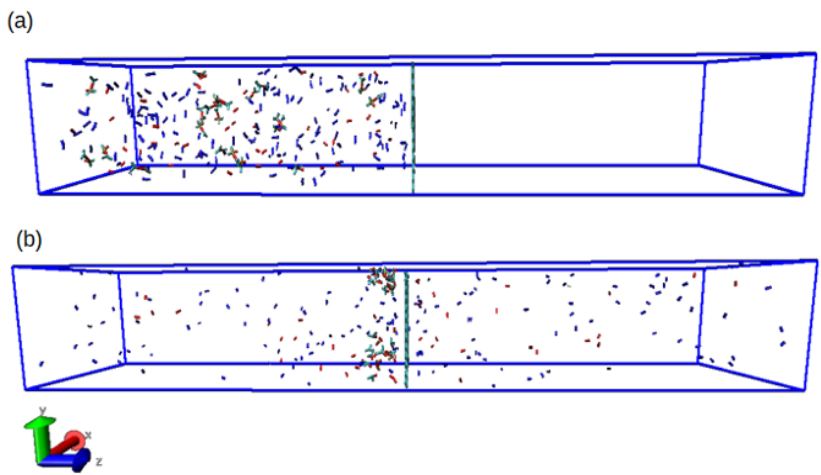

Figure 2. Snapshots of the (a) initial and (b) final configurations for 20 sarin molecules and 100 air molecules (17% of sarin concentration). The final configuration was taken after 10 ns.

The simulations of sarin-air-graphene system were performed on a NVT ensemble, with periodic boundary conditions along $x$ and $y$ directions, and using a Nose-Hoover thermostat [17-18]. The timestep of all simulations were 1 fs and the total time for each simulation was 10 ns. The structures and their properties were calculated from the results of the last 1 ns of simulation. Twenty molecules of sarin were used in all simulations and the quantity of air was changed to control the sarin concentration. For 50, 100 and 200 air molecules, keeping the proportion of nitrogen, oxygen and argon, sarin concentrations of were 28%, 17%, and 9%, respectively. Simulations were repeated for each of the three different sarin concentrations and for each of the following values of temperature: $T = 300$, 500, 700 and 900 K. Figure 2 shows one example of the initial (a) and final (b) configurations for the sarin concentration of 17% (20 sarin molecules and 100 air molecules), for T = 300 K.

**RESULTS AND DISCUSSIONS**

In order to verify the sarin filtration, we counted the number of sarin molecules ($N_s$) that crossed and remained on the initially empty side of the box after 10 ns. The Table I summarizes the results for the simulations with all sarin concentrations (9%, 17% and 28%) and for all temperatures (300 K, 500 K, 700 K and 900 K). It shows that larger the temperature, larger the number of sarin molecules that passed through the graphene hole.

Table I. Number of sarin molecules that is on the initially empty box side after 10 ns by passing through the porous of graphene.

|  | 300 K | 500 K | 700 K | 900 K |
|---|---|---|---|---|
| $N_s$ (9%) | 0 | 0 | 3 | 5 |
| $N_s$ (17%) | 0 | 0 | 2 | 8 |
| $N_s$ (28%) | 0 | 0 | 2 | 13 |

Figure 3 shows the radial distribution function, $g(r)$, between the carbon on graphene and the oxygen atom in the central position of the molecule of sarin, for $T = 300$, 500, 700 and 900 K, and for 17% of sarin concentration. We see that sarin molecules tend to cluster near graphene, being mainly located between 6 and 12 Å. However, for 900 K, the amount of $g(r)$ decreases indicating that some sarin molecules, at this temperature, spread out through the box.

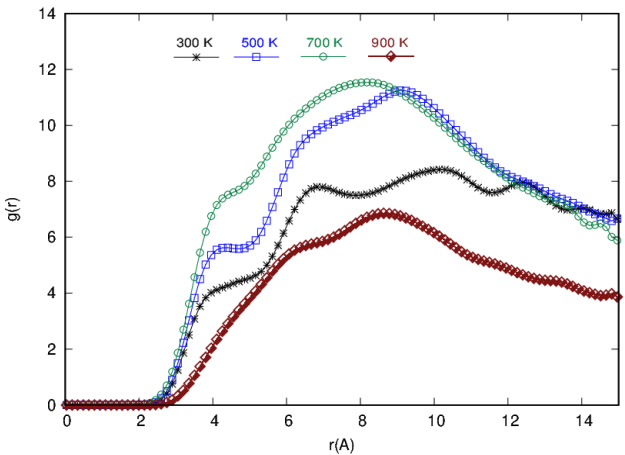

Figure 3. Radial distribution functions of a carbon of graphene (C) and central oxygen atom from sarin (O1s), for different temperatures.

At room temperature, not a single sarin molecule passed through the hole (see Table I), and that sarin gas remain close to graphene. It suggests that there is an attractive force between sarin and graphene. As the net charge of graphene atoms are set to zero, there exists only van der Waals interactions between graphene and sarin. In order to investigate the attraction between them, we calculated the static energy of the porous graphene and one molecule of sarin, as function of the distance between graphene plane and the sarin O1s atom, in three different orientations (we assumed the sarin orientation as the direction formed by the phosphorus-oxygen bond: perpendicular, parallel, and tilted to the plane of the graphene nanosheet).

Figure 4 shows the potential energy curve for the perpendicular orientation. It is possible to verify an energy well of about 1.4 kcal/mol (therefore, larger than the thermal energy, $kT \sim 0.6$ kcal/mol at room temperature). The values of the potential wells are 1.5 kcal/mol and 1.2 kcal/mol for parallel and tilted orientations, respectively. These results are coherent with those of Table I and Figure 3. They show that at 900 K some sarin molecules passed through graphene hole while other sarin molecules spread out through the simulation box. As the attractive potential energy well is about 3 times that of thermal energy at room temperature, at a temperature of about 3 times 300 K (or about 900 K), the molecules get enough thermal energy to escape the potential well. This explains why some molecules pass through the graphene hole or become far from the graphene.

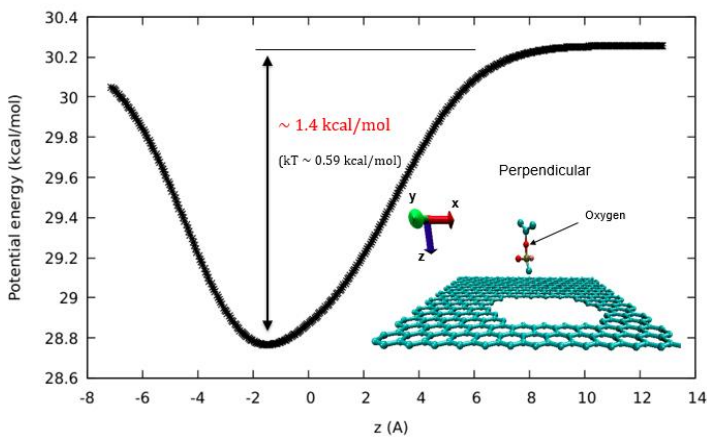

Figure 4. Potential energy vs distance between O1s and the graphene plane, for the sarin molecule perpendicular to graphene. In the details, the estimate of the potential well and the value to the thermal energy at room temperature.

The air permeation through the graphene hole was also evaluated. The distribution of sarin and air molecules along $z$ direction of the simulation box was computed and shown in Figure 5, for 300 K, and for sarin concentrations of 28% (a), 17% (b), and 9% (c). One can see that air molecules spread throughout the box although it prefers to concentrate near graphene and sarin. Again, van der Waals attractive forces might be responsible for the concentration of air close around graphene. But these forces are not enough to prevent the air from passing through the graphene. If, at 300 K, sarin is blocked and air molecules pass through the porous graphene, it shows, at least for the structure simulated here, that the filtering of sarin molecules at 300 K works. It remains to analyse the effect of the hole size.

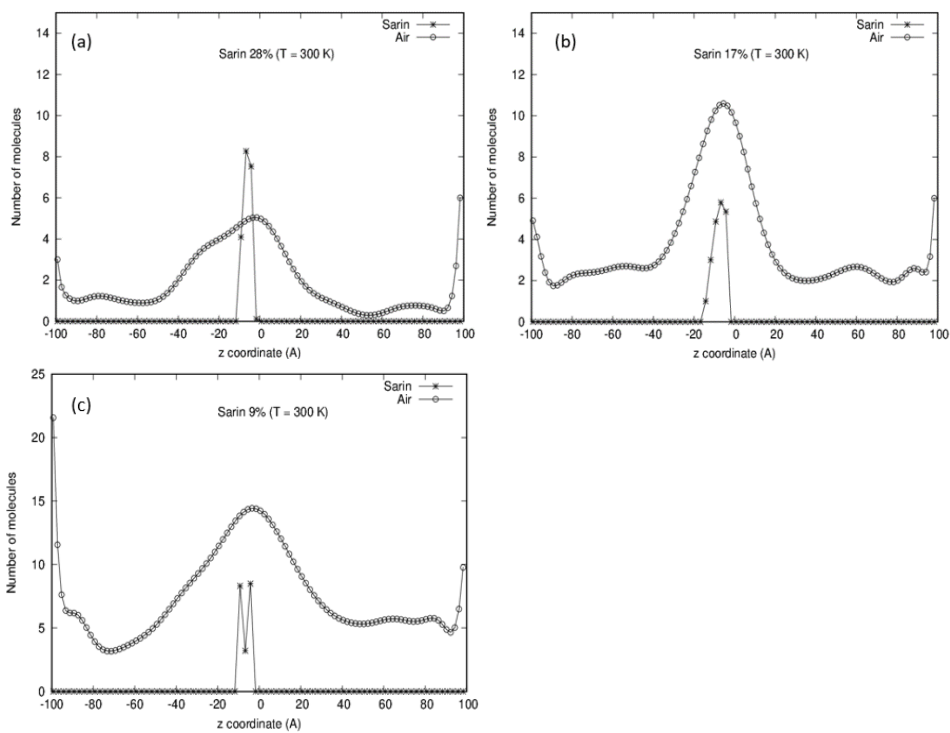

Figure 5. Air and sarin molecules distributions along $z$ direction of the simulation box after 10 ns for $T = 300$ K and for sarin concentrations of 28% (a), 17% (b) and 9% (c). The points represent the average values (over 100 frames taken from the last 1 ns of simulation) of number of molecules at each value of the $z$ coordinate.

In order to evaluate the effect of the hole size on the effectiveness of porous graphene as a sarin filter, we performed the same potential energy calculation between one sarin molecule and a perfect graphene. The result is shown in Figure 6 for the perpendicular orientation of the sarin molecule. The potential well is more than 3 times larger than that for the porous graphene. This clearly indicates that larger the hole size, smaller the potential energy well, smaller the effectiveness of porous graphene as a sarin filter.

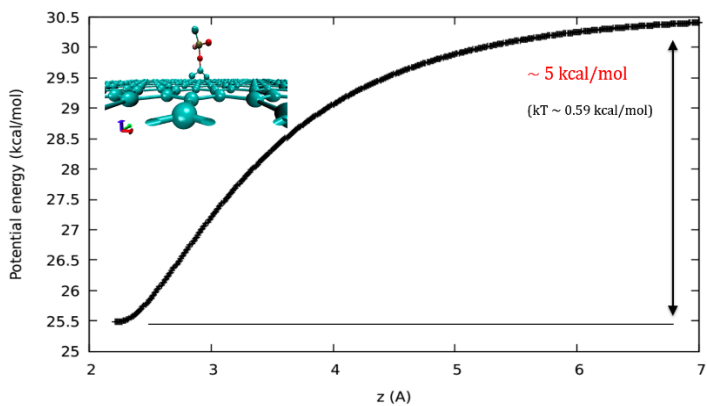

Figure 6. Potential energy vs distance between O1s and the graphene plane, for the sarin molecule perpendicular to a pristine graphene. In the details, the estimate of the potential well and the value to the thermal energy at room temperature.

## CONCLUSIONS

We have performed molecular dynamics simulations of sarin-air-porous graphene systems, in order to investigate the possibility of using graphene as a sarin filter. Our results show that as far as the graphene pores are small enough, of about 1 nm of diameter size, graphene is able to block and hold sarin molecules. As the concentrations of sarin considered in our simulations are higher than the lethal one, our results might be considered promising.

The results form a reference to which the next studies of interaction of sarin and graphene oxide (GO) should be compared.

## ACKNOWLEDGMENTS

MAM acknowledges support from UFSCar and FACENS. AFF is a fellow of the Brazilian Agency CNPq and acknowledges the grant #2018/02992-4 from São Paulo Research Foundation (FAPESP), and the support from FAEPEX/UNICAMP. This research also used the computing resources and assistance of the John David Rogers Computing Center (CCJDR) in the Institute of Physics "Gleb Wataghin", University of Campinas.